\documentclass[english,preprint]{revtex4-1}
\usepackage[latin9]{inputenc}
\usepackage{float}
\usepackage{textcomp}
\usepackage{amstext}
\usepackage{graphicx}

\makeatletter

\newcommand{\lyxmathsym}[1]{\ifmmode\begingroup\def\b@ld{bold}
  \text{\ifx\math@version\b@ld\bfseries\fi#1}\endgroup\else#1\fi}

\providecommand{\tabularnewline}{\\}

\makeatother

\usepackage{babel}
\begin{document}

\title{Photoemission-based microelectronic devices}

\author{Ebrahim Forati$^{1,*}$, Tyler J. Dill$^{2}$,
Andrea R. Tao$^{2}$, and Dan Sievenpiper$^{1,\dagger}$}

\affiliation{$^{1}$ Electrical and Computer Engineering Department, University
of California San Diego, La Jolla, CA 92098}

\affiliation{$^{2}$ Department of NanoEngineering, University of California San
Diego, La Jolla, CA 92098}

\email{dsievenpiper@eng.ucsd.edu}

\email{forati@ieee.org}

\begin{abstract}
The vast majority of modern microelectronic devices rely on carriers within semiconductors due to their integrability. Therefore, the performance of these devices is limited due to natural semiconductor properties such as band gap and electron velocity. Replacing the semiconductor channel in conventional microelectronic devices with a gas or vacuum channel may scale their speed, wavelength, and power beyond what is available today. However, liberating electrons into gas/vacuum in a practical microelectronic device is quite challenging. It often requires heating, applying high voltages, or using lasers with short wavelengths or high powers. Here, we show that the interaction between an engineered resonant surface (metasurface) and a low-power infrared (IR) laser can cause enough photoemission (via electron tunneling) to implement feasible microelectronic devices such as transistors, switches, and modulators. Photoemission-based devices benefit from the advantages of gas-plasma/vacuum electronic devices while preserving the integrability of semiconductor-based devices.
\end{abstract}
\maketitle

\section{Introduction}

In $1906$, the first vacuum-based electronic device, a diode, was
invented by Fleming, and later in $1907$ Lee De Forest introduced the
first vacuum-based amplifier. Low pressure gas was then added to the
vacuum tubes to increase their power handling due to the excess current
generation by the ionized gas. During the $1960$s and $1970$s, vacuum and gas-plasma
electronic devices such as voltage regulators, switches, and modulators
were widely used in RF communication and audio systems. After being
mostly replaced by semiconductor counterparts, due to their integrability,
research on vacuum electronic devices was mostly directed towards high power traveling wave tubes and THz sources. In addition, gas-plasma devices (usually with micro-scale dimensions) have been studied for plasma displays, water treatment, ozone generation, pollution
control, medical treatment, and material processing \cite{singh2014metamaterials,p2011microplasmas,foest2006microplasmas,tendero2006atmospheric}.

On the other hand, further optimization of semiconductor devices is becoming more challenging due to the limitations of the natural properties of semiconductors such as bandgap and electron mobility. For some applications, replacement of semiconductors with substitute  materials may open up new opportunities for scaling characteristics of existing electronic devices such as the speed, power, wavelength, etc.  For instance, vacuum or gas plasma devices benefit from higher mobility of electrons than their semiconductor counterparts. As and example, the electron
mobility under an electric field strength of $10^{3}\,$V/cm in neon
gas (at pressure $100\,$Torr and temperature $300\,$K leading to
the atomic density of $N_{e}=3.2\times10^{18}\,\mathrm{c}\mathrm{m}^{-3}$)
is greater than $10^{4}\,\mathrm{c}\mathrm{m}^{2}\mathrm{V}^{-1}\mathrm{s}^{-1}$.
This mobility is $\sim7$ times larger than the electron mobility
in silicon (Si) at $300\,\mathrm{K}$ which is $\mu_{m}=1350\,\mathrm{c}\mathrm{m}^{2}\mathrm{V}^{-1}\mathrm{s}^{-1}$
\cite{tchertchian2011control,brown2013surface}. The higher mobility
of an electron in gas plasma is mostly due to the lower number density
of atoms (typically 4-6 orders of magnitude lower) compared to semiconductors.
This comparison becomes bolder if we replace gas plasma with vacuum. 
However, issues such as gas plasma ignition (which requires high static voltages or high laser intensities), electrode erosion due to gas atom collisions in plasma,  electron injection into vacuum (typically by thermoionic emission), and the lack of integrability with other (semiconductor) micro-devices have reduced development of  micro-plasma and micro-vacuum devices  to compete with semiconductor microelectronics.  

Here, we propose to use the combination of photoemission (assisted by localized surface plasmon resonances (LSPRs)) and field emission in order to inject electrons into the surrounding space (vacuum or gas) and therefore to realize semiconductor-free microelectronic devices such as switches, transistors, photo-detectors, etc. We show that, by exciting LSPRs, a simple low power diode laser ($mW$ range) along with a small bias voltage ($<10\,V$) can activate a semiconductor-free device. Due to their small dimensions (micro scale) and fabrication method (lift-off process), the proposed photoemission-based devices can also be integrated with semiconductor devices. 

\begin{figure}[t]
\includegraphics[height=2in]{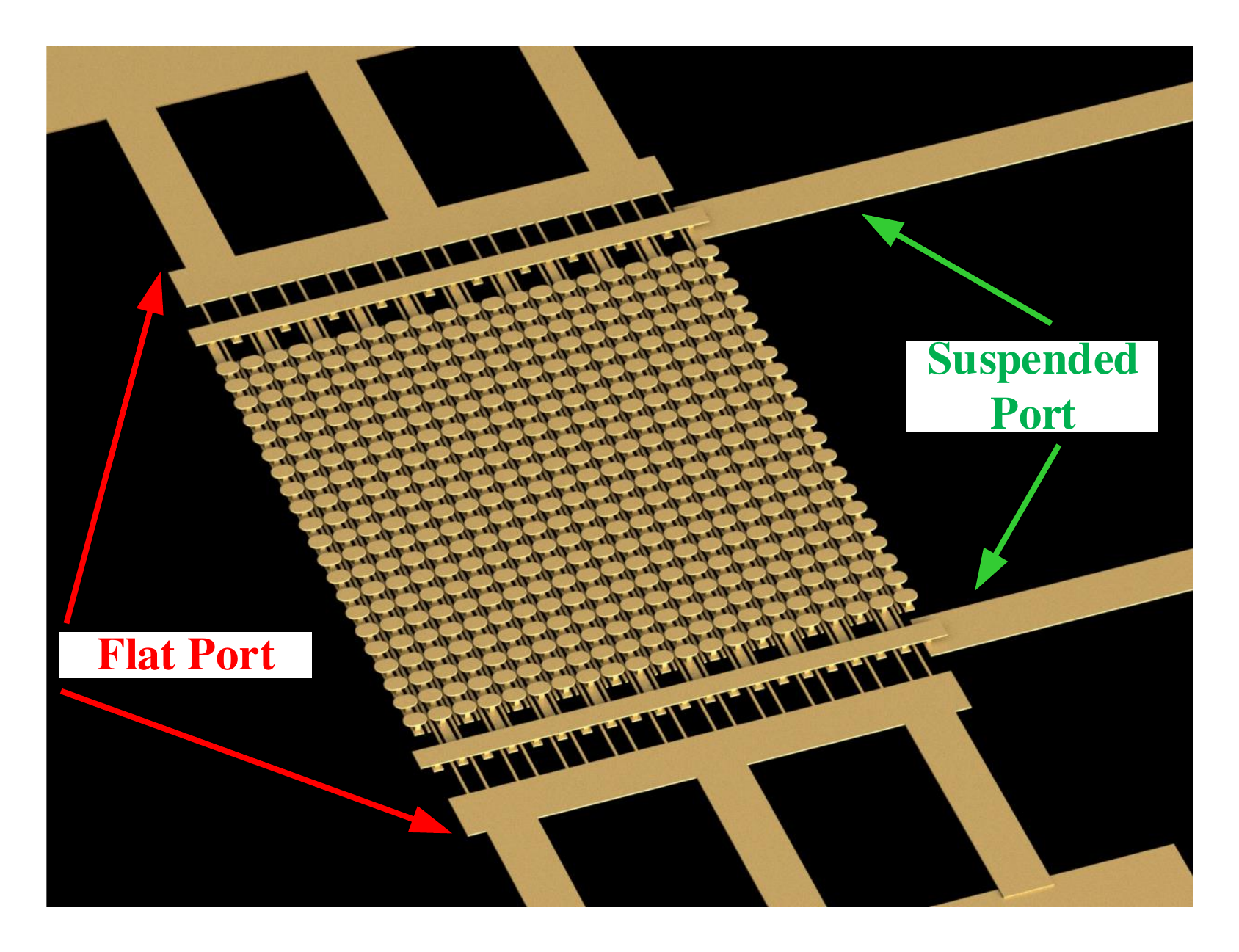}\includegraphics[height=2in]{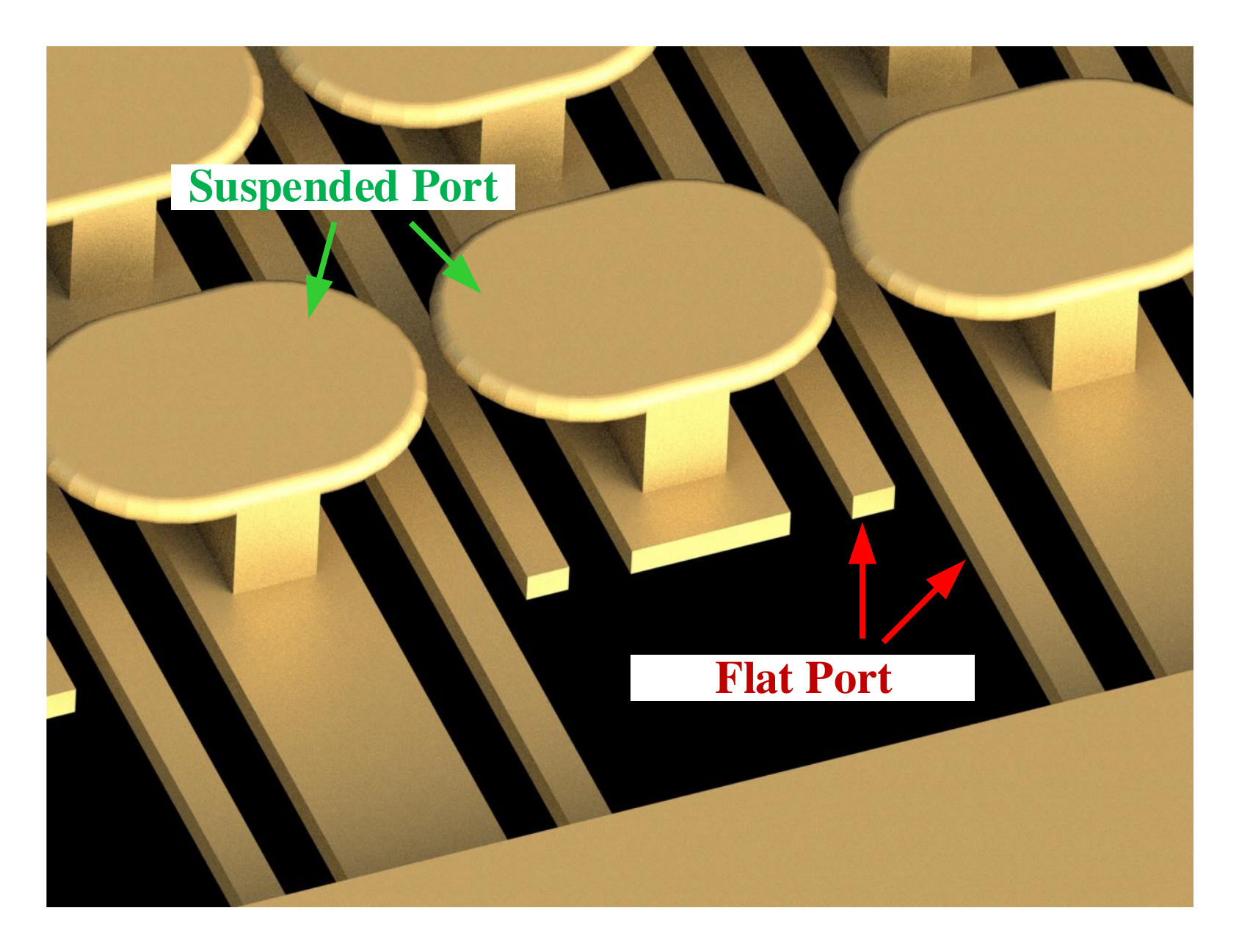}

\caption{The designed photoemission-based device. Biased resonant inclusions under illumination by a wavelength-tuned
CW laser can cause electron emission. The free electrons can be manipulated electrically by proper applied voltages on (nearby) inclusions.}
\end{figure}

The four main mechanisms which free electrons from a material (mostly metals) are: thermoionic emission, the photoelectric effect, electric field emission, and photoemission. 
In a thermoionic emission process, electrons are transferred over the surface potential barrier of the metal (the work function) due to the added thermal energy.  Most of the vacuum electronic devices, including vintage triodes and modern magnetrons, rely on thermoionic emission. However, thermionic emission requires cathode temperatures on the order of 1000 K which makes it infeasible in micro-scale dimensions.  In the photoelectric effect, discovered by Heinrich Hertz in 1887 \cite{mulligan2008heinrich}, photons with higher energy than the material's work function (normally in the ultraviolet range) couple to electrons and emit them over the work function. This was used in both vacuum and gas-filled phototubes, which were eventually sperceded by semiconductor photo-resistors and photo-diodes. Electric field emission, also called Fowler-Nordheim tunneling, is the process whereby electrons tunnel through the lowered work function in the presence of a high electric field (typically static). This has been investigated for realizing cold cathode emitters to replace thermoionic emitters, since they do not require high power to thermally extract electrons from the surface, yet they require bias voltages above $100\,V$ \cite{whaley2009100,cole2014deterministic}. In a photoemission phenomenon, high photon numbers either tunnel electrons through the potential barrier, or transfer them over the barrier (multi-photon absorption) \cite{jouin2014quantum,kruger2011attosecond,sussmann2015field,herink2012field}. Unlike the photoelectric effect, the photon's energy in photoemission is less than the metal work function, and the key factor is the laser-matter interaction (either strong-field or perturbative), caused by the nanolocalized electromagnetic field in the vicinity of metallic structures (such as sharp metallic nanotaperes).  Typically, laser intensities on the order of $TW/cm^{2}$ (in the IR range) are required for photoemission \cite{piglosiewicz2014carrier}. However, photoemission can be greatly enhanced by the excitation of collective electron modes of the metal, called surface plasmon polaritons (SPPs), and laser intensities on the order of  $GW/cm^{2}$ in the mid-IR have been shown to be enough for photoemission \cite{teichmann2015strong}. Nonetheless, the aforementioned electron emission mechanisms can be combined in order to increase the yield. For example, in the so-called thermo-field regime, the cathode temperature is elevated simultaneously with an applied electric field. Also, historically, these emission processes have been mostly studied for vacuum, and well-established theories exist to estimate their current densities. However, the same conclusions can be used at higher pressures for dimensions smaller than the mean free path of electrons (about $\mu m$ in air).

Here, we show that unprecedented laser intensities of around $W/cm^{2}$, e.g. a simple continuous wave (CW) near-IR diode laser, would suffice to trigger the photoemission.  This will be done by exciting LSPRs simultaneously with applying a relatively low static electric field on the order of $10\, V/\mu m$.  The frequency at which LSPRs are supported can be controlled in the design to some extent, which can be seen as another advantage of the photoemission-based devices.  We also show that this combination (electric field emission and photoemission) is a feasible and robust method for controlling the emitted current, which can lead to devices such as transistors, switches, modulators, etc.

We fabricated an engineered micro-surface which supports LSPRs in the near-IR range, and enabled us to apply static voltage between inclusions of the surface. In our scheme, combined photonic and electric excitation of a metallic micro-surface causes electron emission and acceleration into the surrounding space. External electric or magnetic fields can then be applied to guide or manipulate these electrons for different device realizations.
Figure 1 depicts our designed two-port device to study the photoemission simultaneously with an applied static bias. In the desinged device, electron emission occurs at the high electric field spots between the resonant inclusions (due to both LSPR and the static bias). The intensity of the electric field at the hot spots can be controlled both electrically (with static bias) and optically (with the incoming laser). We will show that the two isolated ports in Fig. 1 can couple together due to the free electrons caused by photoemission.

\begin{center}
\begin{figure}
\includegraphics[width=6in]{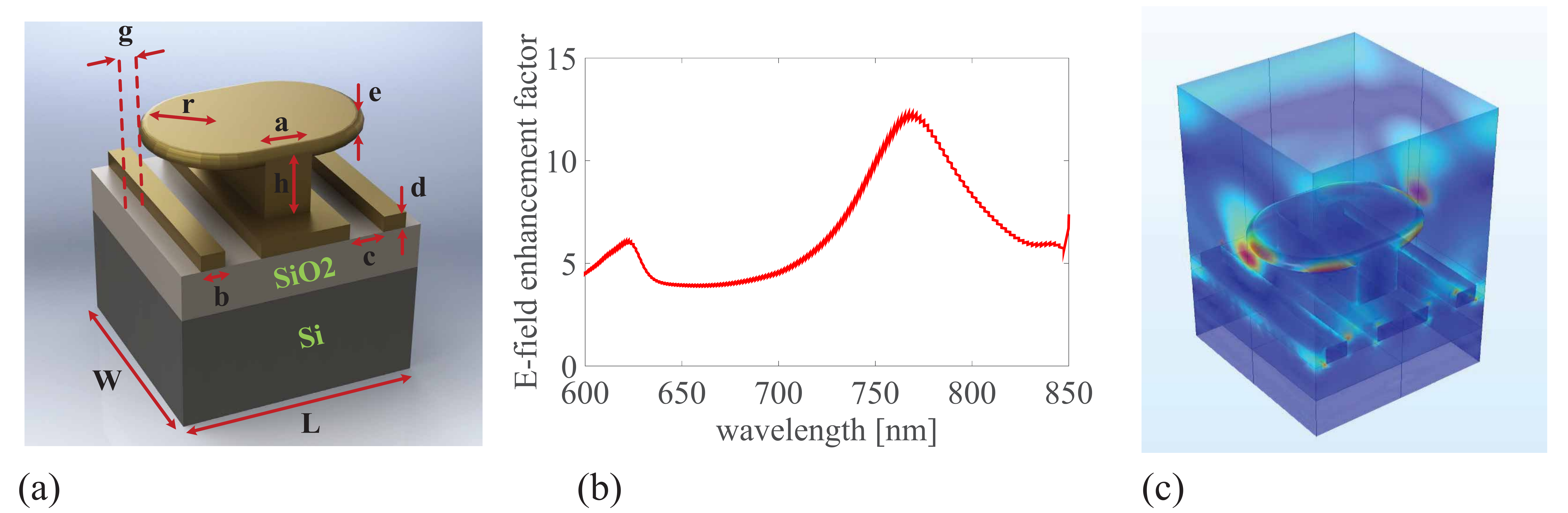}\caption{Unit cell of the resonant surface (a) and its full wave simulated
electric field enhancement (b) and norm distribution (c). The dimensions
are $a=100$, $b=100$, $c=150$, $d=80$, $e=70$, $g=50$, $r=240$,
$L=850$, $W=880$, all in nanometers. The field distribution is at
$\lambda=785\,nm$, and red color represents the highest value.}
\end{figure}

\par\end{center}

\section{Designing the metasurface and results}

The design was inspired by surface-enhanced Raman scattering (SERS)
in which the local electric field is greatly enhanced due to the surface
roughness in order to amplify the Raman response
of biomolecules \cite{chirumamilla20143d,crozier2014plasmonics,brown2013surface}.
The intention in the design of the device was to exploit the distributed
high Q resonance, inherent in certain periodic structures, to dramatically
enhance the absorption of local photons and facilitate photoemission \cite{sievenpiper1999high,yao2013high,eggleston2015optical}.


Figure 2 shows the unit cell of the high Q resonant surface which
we used to electro-optically emit electrons. The unit cell consists
of gold metallic inclusions, vertical gold posts topped with gold
plates, on a silicon (Si) wafer with a layer of silicon dioxide (SiO2)
in between as isolation. Silicon wafers with a layer of SiO2 (typically
between $100-600\,\mathrm{nm}$) are usually used as the substrate
in photo-detection devices. The SiO2 layer is used as an isolator
to minimize the leakage current in the device. Usually a $200\,\mathrm{nm}$
thick layer of SiO2 provides enough isolation \cite{gauthier1995engineering}. The Si wafers used
in our experiments had 1000 $\Omega\mathrm{cm}$ resistivity and the
SiO2 layer was coated on the wafer using plasma sputtering. Full wave
simulation of the unit cell, included in Fig. 2, confirms a resonance
at $\lambda=785\,\mathrm{nm}$ with the electric field enhancement
of about $EF=12$ (defined as the ratio of the maximum to the incident electric field at the gap center) under proper linear polarization (along the mushroom's
length). The field enhancement is due to the localized surface plasmon
resonance supported by gold \cite{maier2007plasmonics,maier2006plasmonic,maier2006effective,yao2009plasmonic,crozier2014plasmonics,wang2006general}.
The resonant mode was optimized so that the enhanced electric field
at resonance (hot spot) is confined to the gap between mushrooms.
As a result, the maximum static electric field (due to the bias), is superimposed
with the laser-induced hot spot. Nonetheless, the flat port (as defined in Fig. 1(b)) also experiences field enhancement of about half of the maximum FE, which will be shown later that is sufficient to emit electrons.

\begin{center}
\begin{figure*}[t]
\centering{}\includegraphics[width=6in]{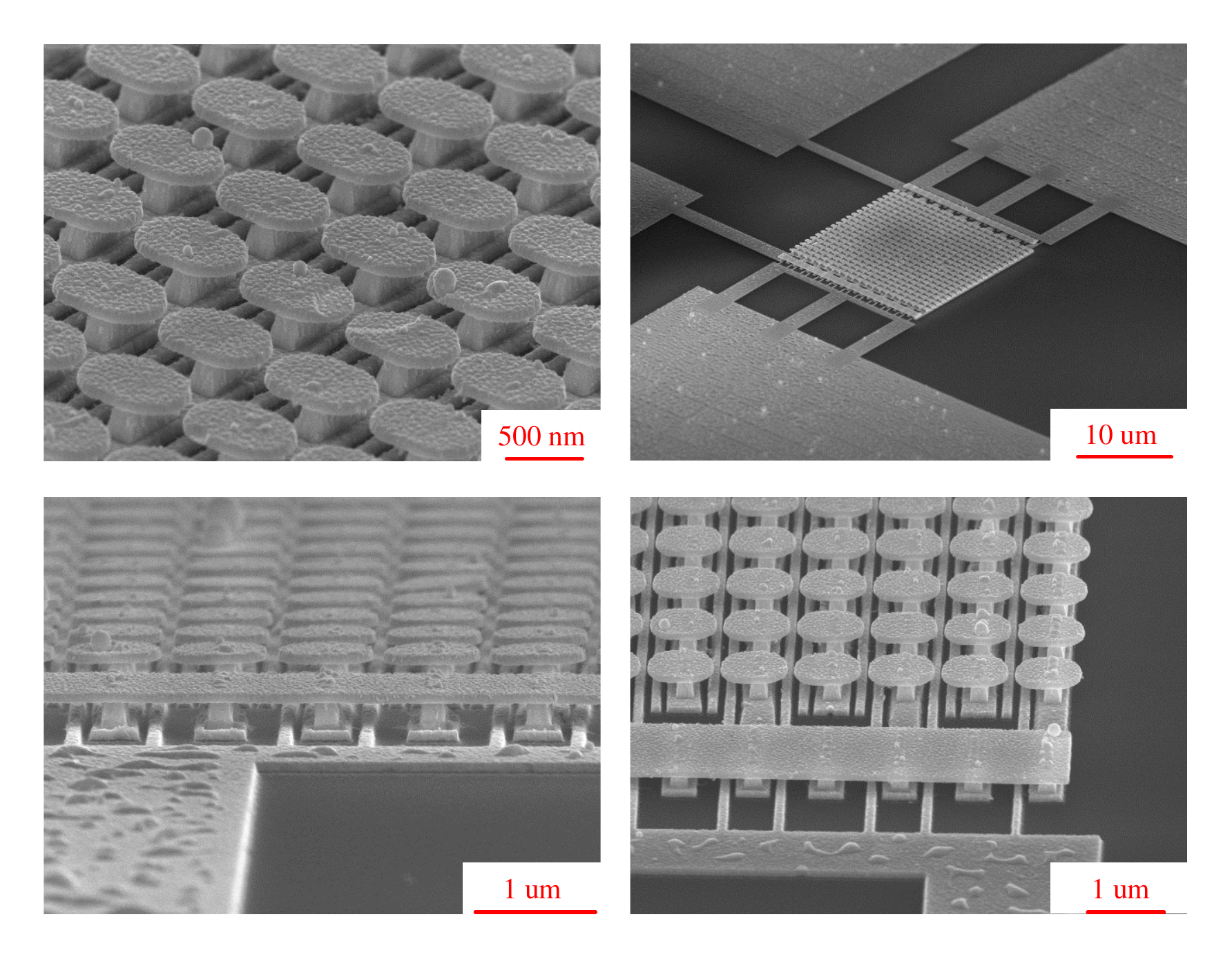}\caption{SEM pictures of the fabricated electron emission-based device. Mushroom rows are to be biased with alternating polarities, using the two airbridges on the side, to form the suspended port. The parallel strips on the substrate, below the mushrooms, form the flat port. }
\end{figure*}

\par\end{center}

The unit cell in Fig. 2 also provides two electrical ports. The
suspended electrical port consists of mushrooms, while the second electrical
port is formed by the gold ribbons on the substrate, as shown in Fig.
1(b), and is called the flat port. With this generic design, we will quantify an important coupling parameter
between the two ports, i.e. transconductance, due to photoemission. Figure 3 shows scattering electron microscopy
(SEM) pictures of the fabricated device
including an array of 21 by 21 unit cells connected to four large
square pads ($250\,\mu m^{2}$) for wire-bonding. In order to form
the suspended  port, so that the static electric field hot spots lay inside the
gap between the mushrooms, we needed to feed every row of mushrooms
with alternate polarities. This was done by placing two air bridges
on the surface sides, and connecting them to the wire bonding square
pads, as clarified in Fig. 3. As a result, after biasing the suspended port,
adjacent mushroom rows will have opposite polarities, similar to an
inter-digital capacitor. The ribbons on the substrate were also connected
to the remaining two square pads, forming the flat port. The surface
was fabricated using a multi-step Ebeam lithography technique, as
discussed in the methods section. 

In order to confirm the high Q resonance of the surface, four different
fabricated samples were characterized using Raman spectroscopy. For
each sample, the electric field enhancement factor was measured at
15 different locations (using a stripe diode with dimensions 1.6$\,\mu$m by 16.95$\,\mu$m). Their averages and standard deviations 
are reported in Table 1. Details of the Raman spectroscopy and field
enhancement determination is reported in the methods section. Based
on the full-wave simulation results, the ratio of the maximum electric
field (at the hot spot) to the average resonance-enhanced electric field on the surface
 is 1.24. Therefore, the maximum FE of surfaces can
be approximated by multiplying the average FEs reported in Table 1
by a factor of 1.24. Based on the results in Table 1, the average
field enhancement of the samples was around $30\times1.24$, which provides substantial photoemission, as will be shown later. Obtaining
very large enhancement factors is challenging due to metallic loss,
as discussed in Refs. \cite{khurgin2015deal,west2010searching}. To
verify resonance, the FE of one of a sample was measured off resonance
(at $\lambda=633\,\mathrm{nm}$) which was almost half of the resonant
FE, as reported in Table 1. This is consistent with the full wave
simulation result, shown in Fig. 2 as well. Reasons for observing higher
experimental FEs than simulation results include surface roughness,
chemical enhancement, and non-linearity of gold polarizability. 

\begin{center}
\begin{table}
\begin{tabular}{|c|c|c|c|c|}
\hline 
Sample \# & $\begin{array}{c}
\lambda=633\,\mathrm{nm}\\
FE_{\mathrm{ave}}
\end{array}$ & $\begin{array}{c}
\lambda=633\,\mathrm{nm}\\
SD
\end{array}$ & $\begin{array}{c}
\lambda=785\,\mathrm{nm}\\
FE_{\mathrm{ave}}
\end{array}$ & $\begin{array}{c}
\lambda=785\,\mathrm{nm}\\
SD
\end{array}$\tabularnewline
\hline 
\hline 
1 & - & - & 25.80  & 0.90\tabularnewline
\hline 
2 & - & - & 23.00  & 0.97\tabularnewline
\hline 
3 & - & - & 24.78  & 0.98\tabularnewline
\hline 
4 & 13.10 & 1.38 & 27.51 & 1.38\tabularnewline
\hline 

\end{tabular}
\caption{The standard deviation (SD) and the average of the the field enhancement on samples' area.}
\end{table}
\par\end{center}

The fabricated samples were then installed and wire bonded inside
standard dual in-line packages, as shown in the supporting information
(SI). 

As the first experiment, the conductivity change of the suspended and flat ports are measured and reported in Fig. 4(a). It is evident that
the optical port illumination varies the conductivity of the suspended and flat ports
sufficiently to realize ON and OFF states, i.e. the structure performs as an optical switch. The change in the conductivity
is caused by the photoemitted electrons from the resonant inclusions on the surface, combined with static field emission at higher bias voltages.  Based on the current versus voltage (I-V) curves on Fig. 4(a), the conductivity of the suspended port increases by a factor of 10 after laser illumination (with the 10 volts bias). Due to the symmetry, we would expect that $V_{f}=0$ leads to $I_{f}=0$ even with  laser illumination (as electron emission from the port's inclusions are symmetrical). However, $I_{f}$ has some negative value with laser illumination which we suspect is due to some asymmetry in the flat port fabrication. The magnitude of this photoemission current increases from $100\, nA$ to $800\, nA$ as the laser intensity increases from $5\,W/cm^{2}$ to $S=40\,W/cm^{2}$. This can also explain the asymmetry in the I-V curve of the flat port in Fig. 4(a).
The laser wavelength and intensity in this experiment were $\lambda=785\,\mathrm{nm}$ and $S=5\,W/cm^{2}$, respectively,  which are easily achievable with a low cost diode laser.  Throughout the experiments, we set the pressure in the $10^{-4}\,\mathrm{Torr}$ range, in order to prevent any gas plasma formation around the device due to the static bias or laser illumination. That is, the device is placed inside a vacuum chamber with some electrical feedthroughs and optical view ports (see SI). This ensures that photoemission is the prevalent mechanism in the device.  However, all of the results reported in this manuscript were also  observed at atmospheric pressure (with slight differences).  More specifically, at lower pressures, the conductivity due to the incoming photons is slightly higher than at air pressure, and the I-V curves are smoother. This is consistent with our expectations at lower pressures due to the reduced scattering of emitted electrons by gas atoms.
Figure 4(a) also shows that the I-V curve of the flat port has asymmetry (versus the static bias polarity) which we suspect is associated with some physical asymmetry in the fabricated  flat port.   Figure 4 also includes the responsivity of the suspended port. As we expected, electron photoemission has a peak at $\lambda=785\,\mathrm{nm}$ which is consistent with the full-wave simulated result shown in Fig. 2(b).  Based on Fig. 4(b), the current yield slope decreases significantly at laser intensities around $20\,W/cm^2$ which indicates switching from the perturbative (multi-photon emission) to the strong-field light-matter interaction (tunneling regime). In other words, the laser intensity of $20\,W/cm^2$, in our design, creates pondermotive energy of electrons comparable with the gold work function (i.e. Keldysh parameter is around unity) \cite{teichmann2015strong}. This typically requires laser intensities above $TW/cm^2$ in the absence of any field enhancement caused by the laser-matter interaction. In ref. \cite{teichmann2015strong}, it is shown that a $1 GW/cm^2$ laser intensity suffices for photoemission if SPPs are excited. Here, we observe that LSPRs (which typically provide higher FE than traveling SPPs) along with a few $V/\mu m$ static electric field lower the laser intensity requirement for photoemission to the $W/cm^2$ range.  It is worth mentioning that the emitted electrons through LSPR-enhanced photoemission are much more energetic than a conventional photoemission, which has been interpreted in terms of pondermotive acceleration \cite{kupersztych2001ponderomotive,dombi2013ultrafast,brongersma2015plasmon}. It comes from the fact that localized electric fields of LSPRs are tightly bounded to the metal surface, on the orders much smaller than the quiver amplitude of an electron. This leads to electrons traveling far from the metal surface before reversing direction during the second half-period of the laser temporal oscillation.

\begin{figure}
\includegraphics[width=6.5in]{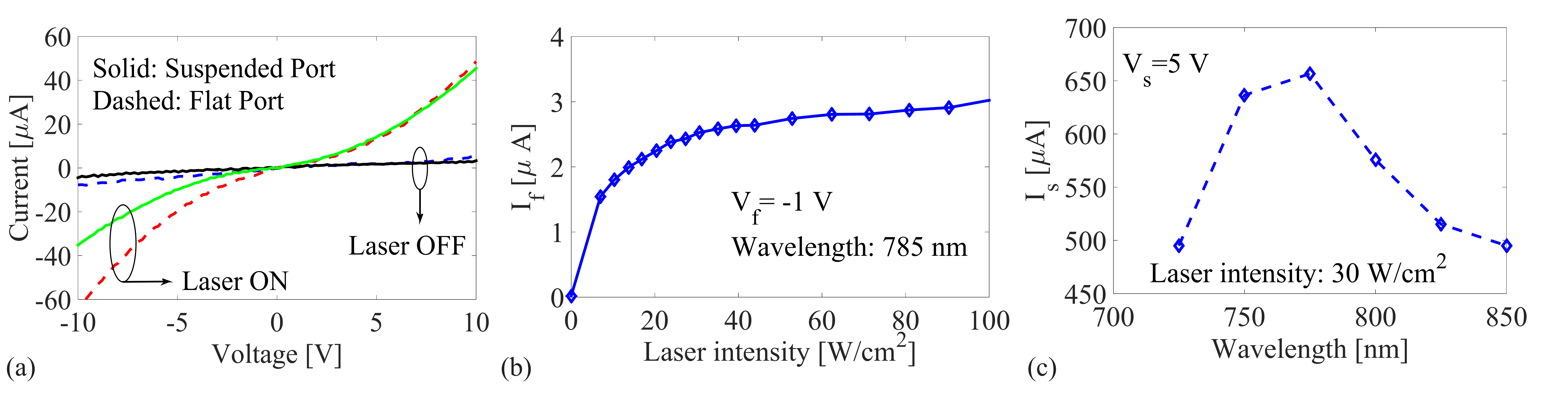}\caption{Individual port responses. a) I-V curves of the suspended/flat port as the flat/suspended port is open-circuited ($S=5\,W/cm^{2}$), b) responsivity of the flat port, c) frequency dependence of the suspended port's response. }
\end{figure}

In all of the experiments of Fig. 4, one port is left open-circuited while measuring the other port. Also, except in Fig. 4(c), throughout this paper the wavelength is set to be $\lambda=785\,\mathrm{nm}$.          

The emitted electrons can be manipulated by external electric or magnetic fields. In our design, this can be studied by measuring the mutual response between the suspended and flat ports, as summarized in Fig. 5. The I-V curve of the flat port with different applied voltages on the suspended port are shown in Fig. 5(a). Since $I_{f}$ is created by the electron emission due to both photoemission and electric field emission, applying a bias voltage on the suspended port with similar or opposite polarity as the flat port increases or decreases the photoemitted current, respectively. For example, without any bias voltage on the suspended port, and with $V_{f}=10\, V$,  the laser illumination changes the flat port's conductivity by a factor of 10 (based on Fig. 4(a)). This conductivity change factor increases to 30 or decreases to  2, with applied bias voltages of +10 and -10  on the suspended port, respectively. Fig. 5(a) demonstrates successful control of the flat port by both optical and suspended ports, which resembles a (semiconductor-free) transistor.

In order to quantify the mutual response of the two ports, the flat port was short-circuited and its current was measured as a function of both $V_{s}$ and the laser power intensity, as shown in Fig. 5(b). We may approximate the photoemission and field emission contributions to the current generation from Fig. 5(b). It is evident that increasing the laser intensity is decreasing $I_{f}$ regardless of the $V_{s}$ polarity. This suggests us that the photoemission current always has negative values in our measurements (this is consistent with the fact that electrons always leave the metal due to photoemission). On the other hand, direction of the field emission current depends on the polarity of the applied static voltage. As a result, the net current in the flat port is the sum or difference of photoemission and electric field emission currents for positive or negative $V_{s}$, receptively. For instance, with the laser intensity of $S=40\,W/cm^{2}$, applying positive or negative 10 volts on the suspended port induces $-60\mu A$ or $10\mu A$ on the flat port, respectively. This leads us to the conclusion that, for the specific laser intensity and bias voltage, the photoemission and field emission currents are  $-25\mu A$ and $+35\mu A$, respectively.   Simple calculations of the generated photoemission current and the photon energy at $\lambda=785\, nm$ shows that photon to electron conversion rate is approximately 5 percent.

Another important information which Fig. 5(b) carries is the rate of the change in $I_{f}$ as $V_{s}$ varies (i.e. the slope of the curves in Fig. 5(b)). This parameter which can be considered as  the small-signal transconductance of the device, is drawn in Fig. 5(c). Although the device is not optimally designed for this purpose, Fig. 5(c) implies an electro-optical transistor whose transconductance can be controlled both with the bias voltage (the horizontal axis) and the photon number. For instance, the bias voltage of 8 volts and laser intensity of $40\,W/cm^{2}$ leads to the transconductance of $10\,\mu S$. Figure 5(d) shows the generated $I_{f}$  as the input ($V_{s}$) of $1\,V_{p-p}$ biased on $+8\,V$ is applied, with and without the laser illumination ($40\,W/cm^{2}$). 
Note that the flat port for Figs. 5(b-d) was short-circuited in order to solely study the coupled energy from the suspended port to the flat port.

\begin{figure}
\includegraphics[width=6in]{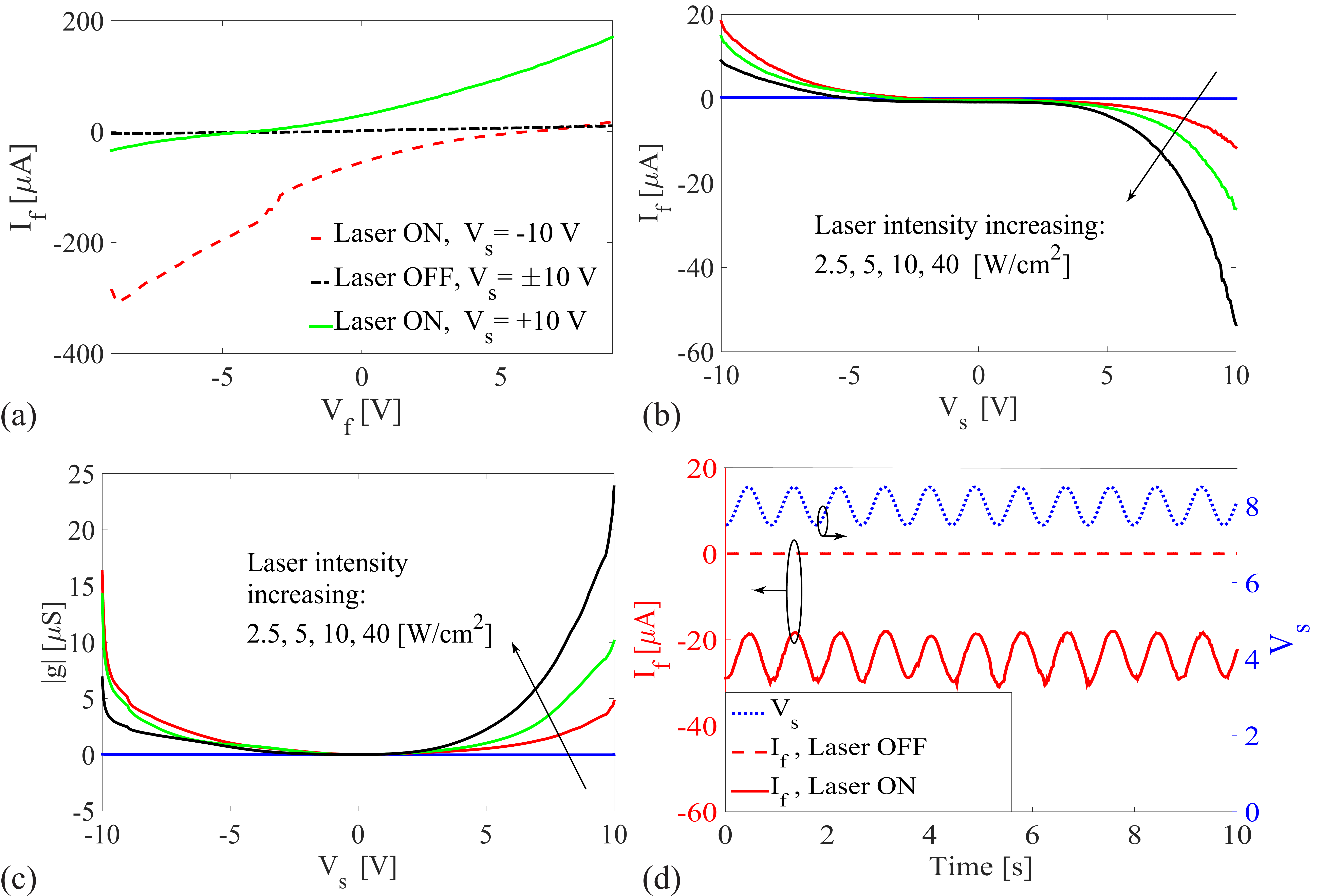}\caption{The mutual response between the two ports. a)  I-V curves of the flat port as the suspended port is biased with different voltages ($S=5\,W/cm^{2}$) , b) the induced current on the flat port as $V_{f}$ is fixed at zero and $V_{s}$ varies, c) the small-signal transconductance of the device ($V_{f}=0$), d) the induced sinusoidal current on the flat port due to the applied sinusoidal voltage on the suspended port ($V_{f}=0$, $S=40\,W/cm^{2}$).   }
\end{figure}

The optical port in our studied device provides complete electrical isolation. Moreover, the electron-emitting surface is highly scalable and therefore is potentially capable of handling
high power. These devices could also be used as photodetectors which
can be tuned to a range of frequencies by adjusting the geometry of the surface.
It can even be designed so that different frequencies resonate with
different regions of the surface, providing a highly sensitive yet
broadband response. 



The bandgap of SiO2 is larger than the photon energy at $\lambda=785\,\mathrm{nm}$ and therefore its resistance does not change with the laser illumination unless the laser causes a change in the temperature, which does not happen during our experiments (we did not use higher laser intensities for this reason). However,  the bandgap of silicon is smaller than the photon energy at
$\lambda=785\,\mathrm{nm}$, and the laser can change its conductivity. If there is a leakage current
through the SiO2 layer and the silicon substrate, some portion of the conductivity change in our experiments could be  due to the silicon contribution. We performed a few experiments
to gather enough evidence that the photoemission/field emission is
dominant in our device. As the simplest experiment, the same I-V curves in Fig.
8 are measured by setting the illumination wavelength at $\lambda=1050\,\mathrm{nm}$,
at which photons do not have enough energy for coupling to electrons in silicon.
The results showed a strong conductivity change (at $\lambda=1050\,\mathrm{nm}$, the resonant structure still has enough FE factor to emit electrons). This was a strong evidence
that silicon absorption was not important at $785\,\mathrm{nm}$
either. As another experiment, the resonant surface was removed from
the design, and a device was fabricated consisting of only the wire
bonding square pads. Assuming that the pads are smooth enough to prevent
field enhancement, the only conductivity contribution in this device
is silicon absorption and temperature rise. The I-V curve of this
device showed a negligible change in the conductivity compared to the photoemission-based device. Moreover, several flat non-resonant devices were fabricated on glass substrate, as a good
insulator, and none of them responded to the illuminating laser, as discussed in the SI.

\section{Conclusion}

We showed that photoemission enhanced by LSPRs and combined with electric field emission can liberate electrons from gold with the unprecedented laser intensities of $W/cm^2$. The fact that low bias voltages (under 10 volts) and low power (a few mW) IR lasers can initiate and control the electron emission is very promising to design semiconductor-free devices with new opportunities to scale their capabilities (such as speed, power handling, etc.) to beyond what is limited by natural properties of today's semiconductors. The substrate in our design only supported the metallic structure and was not involved in the electric current flow. Events that would damage an ordinary semiconductor device (e.g. over-voltage or radiation) would have little effect on a photoemission-based device.

\section{Method}

A commercial finite element method code (HFSS) was used to simulate
the unit cell, and Johnson-Christy model was adopted for gold. In
order to simplify the simulations, mushrooms were set to have a smooth
surface and rounded corners. This ensured that we avoided non-localities
in the gold model and calculation singularities at the sharp corners
\cite{toscano2012surface}. As a result, the simulated field enhancement
is a lower limit and, as will be shown later, the measured field enhancement
will be larger. The measurement setup includes a tunable Ti:Sapphire
laser pumped with a $10\,\mathrm{W}$ green semiconductor laser. The
output laser beam was passed through two beam samplers for wavelength
and power measurements. An Ocean optics spectrometer and a silicon
photo-detector were used for wavelength and power measurements, respectively.
The laser beam was sent into a vacuum chamber through a view port
using a few optomechanics. The vacuum chamber was equipped with a
vacuum pump, Ar gas inlet, pressure gauge, electrical feedthroughs,
and a customized imaging system observing the device from outside
of the chamber. The fabricated devices were installed in standard
dual in-line packages with a small piece of carbon conductive tape
(typically used in SEM) and wire bonded using a ball bonder. Two source-meters
(Keithley 2400 and 2410) with a common ground were used for full characterization
of the three port devices. In order to measure the I-V curves, the
vacuum chamber was pumped down to $10^{-4}\,$Torr (our equipment limit). The negative
electrode of the two sourcemeteres were connected to the optical table
(ground), therefore two terminals of the device's ports were essentially
connected.

\subsection{Fabrication}

A three-layers recipe was developed and optimized to perform the fabrication
of clean mushrooms and air bridges. The first layer consisted of the
gold ribbons on the substrate, the second layer included vias, and
the third layer comprised the mushroom caps. After cleaning the wafer
with acetone, $180\,\mathrm{nm}$ SiO2 was deposited on the wafer
using plasma sputtering. Then, the first layer was patterned and fabricated
using Ebeam lithography and Ebeam evaporation ($70\,\mathrm{nm}$
Au on top of $10\,\mathrm{nm}$ Cr as the adhesion layer). Similarly,
the second layer (vias) was fabricated using Ebeam lithography and
Ebeam evaporation ($250\,\mathrm{nm}$ Au). In order to fabricate
the third layer (mushroom caps), photoresist (AZ1505) was spin-coated
on the sample and was ashed with oxygen plasma down to the thickness
of $200\,\mathrm{nm}$, so that the tip of the vias were exposed.
Then, a few nanometers of chromium was sputter coated on the photoresist
to prevent it from mixing with the Ebeam resist. Next, Ebeam resist
was coated on the sample (without any soft baking) and was patterned
using Ebeam lithography. The samples were then ready after metallization
($70\,\mathrm{nm}$ Au), lift off using acetone, chromium plasma etching,
and oxygen plasma cleaning.

\subsection{Raman spectroscopy/ Experimental Determination of Field Enhancement}

Experimental FEs were determined by comparing the enhanced spectra
of thiophenol, a common SERS marker, to bulk Raman measurements and
then dividing by the respective number of excited molecules. Because
FE is dependent on $(E/E_{0})^{4}$, we can approximate the average
field enhancement over the device with this method. 

Thiophenol forms a self-assembled monolayer (SAM) on gold surfaces.
We performed an overnight thiophenol vapor phase deposition on the
device. Excess thiophenol was removed by placing the device under
vacuum for \textgreater{}2 hours. Raman measurements were conducted
on both bulk thiophenol and on the thiophenol monolayer coating the
gold surface of the device. Both sets of measurements were carried
out using the same measurement configuration. All data was collected
using either a $785\,$nm diode laser, or a $633\,$nm HeNe laser,
at powers of $<1\,$mW, to ensure no desorption of the monolayer or
morphological changes to the gold structure. The FE was then calculated
using 

\[
EF=(\frac{I_{\mathrm{SERS}}}{I_{\mathrm{Raman}}})(\frac{N_{\mathrm{Raman}}}{N_{\mathrm{SERS}}})
\]
in which $I$ is the measured bulk Raman or SERS Intensity, and N
is the number of molecules from which the Raman signal originates.
$N_{\mathrm{Raman}}$ was calculated using the density and molecular
weight of bulk thiophenol along with laser focal volume. $N_{\mathrm{SERS}}$
was calculated from the gold structure area, multiplied by the literature
packing value for thiophenol SAMs of $6.8\,\mathrm{molecules}/\mathrm{nm}^{2}$.
The laser spot size was calculated using the scanning knife-edge method.
The laser spot was scanned over a cleaved Si wafer edge in both X
and Y directions and the $520\,\mathrm{cm}^{-1}$ peak intensity was
recorded over the length of the scan. The plots were fitted to error
functions and the Gaussian beam waists derived. Focal Depth was calculated
by translating the Si along the z-axis, with the focal plane in the
center. This was fitted to a Gaussian and the focal depth was taken
as the integral (-inf, inf) of the fit. Field enhancement  was calculated
using the $999\,\mathrm{cm^{-1}}$ peak because it displays low orientational
dependence on intensity, and is therefore less effected by molecular
reordering on a metal surface. In addition it displays the highest
bulk Raman signal and so gives us the most conservative FE calculation.
Standard deviations were determined with measurements at \textgreater{}15
random points over the device surface.

\section*{Supplementary Information} is available in the online version of the paper.

\section*{Acknowledgements}

The authors thank Shiva Piltan, and UC San Diego nanofabrication facility staff including
Sean Parks, Larry Grissom, Ryan Anderson, Ivan Harris, and Xuekun
Lu for the helpful discussions, and especially Maribel Montero for
performing Ebeam lithography exposures. This work was funded by Defense Advanced Research Projects Agency
(DARPA) through grant N00014-13-1-0618 and ONR DURIP through grant N00014-13-1-0655.

\section*{Author contributions}

Dan Sievenpiper proposed the idea and supervised the study. Ebrahim
Forati conceived and conducted the fabrication and experiments. Tyler
Dill performed and Andrea Tao supervised the Ramon spectroscopy measurements.
Ebrahim Forati and Dan Sievenpiper analyzed the results. Ebrahim Forati
wrote the manuscript. All authors reviewed the manuscript.

\section*{Author Information} Reprints and permissions information is available at
www.nature.com/reprints. The authors declare no competing financial
interests. Readers are welcome to comment on the online version of the paper.
Correspondence and requests for materials should be addressed to Ebrahim Forati (forati@ieee.org) and Dan Sievenpiper (dsievenpiper@ucsd.edu).

\bibliographystyle{apsrev4-1}
\bibliography{Nature_ref}

\end{document}